\documentclass[fleqn,10pt]{wlscirep}
\usepackage[T1]{fontenc}
\usepackage{graphicx}% Include figure files
\usepackage{dcolumn}% Align table columns on decimal point
\usepackage{bm}% bold math
\usepackage{mathrsfs}
\usepackage[utf8]{inputenc}
\usepackage{textcomp}
\newcommand{\ii}{i} %MM: Defining a command for the imaginary unit so that it is easy to change the definition later if needed. I checked from NE and they use Italics i at least in the latest paper.

\title{A low-noise on-chip coherent microwave source}

\author[1,2,*]{Chengyu Yan}
\author[3,4]{Juha Hassel}
\author[4]{Visa Vesterinen}
\author[1,5]{Jinli Zhang}
\author[1]{Joni Ikonen}
\author[4]{Leif Gr\"{o}nberg}
\author[1,3]{Jan Goetz}
\author[1,4,*]{Mikko M\"{o}tt\"{o}nen}

\affil[1]{QCD Labs, QTF Centre of Excellence, Department of Applied Physics, Aalto University,P.O. Box 13500, FI-00076 Aalto, Finland}
\affil[2]{MOE Key Laboratory of Fundamental Physical Quantities Measurement $\&$ Hubei Key Laboratory of Gravitation and Quantum Physics, PGMF and School of Physics, Huazhong University of Science and Technology, Wuhan 430074, P. R. China}
\affil[3]{IQM,  Keilaranta 19, 02150 Espoo, Finland}
\affil[4]{QTF Centre of Excellence, VTT Technical Research Centre of Finland Ltd,P.O. Box 1000, FI-02044 VTT, Finland}
\affil[5]{College of Electronic Information and Optical Engineering, Nankai University, Tianjin, 300071, China}

\affil[*]{corresponding author(s): Chengyu Yan(chengyu$\_$yan@hust.edu.cn), Mikko M\"{o}tt\"{o}nen(mikko.mottonen@aalto.fi)}

\begin{abstract}
The scaleup of quantum computers operating in the microwave domain requires advanced control electronics, and the use of integrated components that operate at the temperature of the quantum devices is potentially beneficial. However, such an approach requires ultra-low power dissipation and high signal quality in order to ensure quantum-coherent operations. Here, we report an on-chip device that is based on a Josephson junction coupled to a spiral resonator and is capable of coherent continuous-wave microwave emission. We show that characteristics of the device accurately follow a theory based on a perturbative treatment of the capacitively shunted Josephson junction as a gain element. The infidelity of typical quantum gate operations due to the phase noise of this cryogenic 25-pW microwave source is less than 0.1\% up to 10-ms evolution times, which is below the infidelity caused by dephasing in state-of-the-art superconducting qubits. Together with future cryogenic amplitude and phase modulation techniques, our approach may lead to scalable cryogenic control systems for quantum processors.

\end{abstract}

\begin{document}                              
\flushbottom
\maketitle

\thispagestyle{empty}

Quantum computing has developed rapidly in the last decade using a range of different physical systems\cite{HSJ08,SHO09,MDL14,NRK18,SAM18,AAB19}. For example, semiconductor and superconductor-based qubits with frequencies in the microwave regime have been studied extensively\cite{HSJ08,SHO09,NFB10,RBT13}. In such systems, the control of a large quantum processor is typically implemented by channelling a sequence of microwave pulses to the qubits operating at many different frequencies. This can be achieved using conventional room temperature electronics. However, the approach requires a large number of broad-band connections scaling linearly with the number of qubits, to transmit signals from room temperature to the refrigerator that hosts the quantum processor at temperatures typically below 100 mK. When scaling-up of these quantum-computing systems, the heavily attenuated bundles of coaxial microwave cables will determine much of the system-level requirements, including the cooling power and the physical size of the refrigerator\cite{KSK19}. Cable lengths also lead to latencies and limitations in, for example, quantum error correction\cite{KNL97,CSH96}.

Cryogenic integrated control electronics can potentially overcome these challenges. Independent of how the cryogenic control electronics are realised, any viable approach will need stable microwave sources integrated in the relevant operating environment to, for example, create the carrier frequencies for the modulated pulses\cite{BOB16}. Josephson-junction-based sources have previously been studied in the field of radio astronomy as local oscillators for receivers in the millimetre and submillimetre wave bands\cite{VSP02}. However, for typical solid-state quantum information processing applications operating in the sub-20-GHz band, the specific requirements are stringent. In particular, an on-chip microwave source should exhibit a very low power dissipation, long coherence time, and low noise. Some of these properties have previously been explored with prototype devices based on quantum dots\cite{LSG15,LHS17,LSE17} and Josephson junctions\cite{CLA14,CBR17} embedded in resonators. Similar designs can also find applications in generating non-classical radiation\cite{GBA19,RPD19}. However, detailed design guidelines for specific applications are generally lacking and it remains unclear whether the signal quality of these systems will be sufficient for high-fidelity qubit operations. 

In this Article, we report an on-chip coherent microwave source based on a Josephson junction strongly coupled to a spiral resonator. We develop a quantitative theoretical model for the resonator-coupled Josephson junction which operates in a particular parameter regime and can yield a stable and coherent continuous-wave microwave output. Our theory is verified in experiments which also provide relevant system parameters, including the output power and the microwave generation efficiency. We also confirm the applicability of our source for quantum-coherent operations by measuring the phase noise of the oscillator output, and provide the total phase noise spectrum up to large offset frequencies and evaluate its impact on the dephasing of an ideal qubit.

We analyse, in particular, the source-induced infidelity of the identity operation and of the NOT gate for an ideal qubit, and conclude that they lie below 0.1\% up to a 10-ms evolution. We achieve this performance metric with a total cryogenic power consumption of 200 pW---which is compatible with the millikelvin environment---and a generation efficiency of about 15\%. We thus suggest that the source is compatible for driving qubits in schemes, in which it is combined with additional amplitude and phase modulation components. Due to the high output power, our on-chip source may also be of potential use in other applications, such as writing and retrieving quantum information encoded in spin ensembles\cite{GJK14} and pumping a microwave-to-optical photon converter\cite{GMS20}.

%\section{Theory and design}
\vspace*{10pt}
\noindent \textbf{Theory and design} 

\noindent Our oscillators are realized as capacitively shunted Josephson junctions (C-shunt JJs) coupled to a resonator. Under certain conditions described below, the phase dynamics of a C-shunt JJ locks to a radiation field that couples to it\cite{HGH06}. The phase-locked and biased junction generates power at the corresponding angular frequency $\omega$. 

Our devices reside in the parameter regime corresponding to relatively high Josephson coupling energies in the range of $E_\textrm{J}/h\gtrsim 1$~THz, where $h$ is the Planck constant. Furthermore, it is known that the required harmonic phase-locking conditions are favored in the case\cite{KAU81} $\omega \gg \omega_\textrm{p}$, i.e., the generated angular frequency $\omega$ sufficiently exceeds the plasma angular frequency of the junction $\omega_\textrm{p} = \sqrt{8E_\textrm{J}E_\textrm{c}}$, where $E_\textrm{c}$ is the charging energy of the junction. The plasma frequencies of bare junctions are typically much higher than our frequency range of interest, 1--10~GHz, which motivates us to decrease $\omega_\textrm{p}$ with a capacitive shunt. To this end, we use an additional parallel-plate capacitor with the effective charging energy $E_\textrm{c} = e^2/(2C_\textrm{s})\approx h\times$ 100--600~kHz, where $e$ is the elementary charge, or equivalently a shunt capacitance $C_\textrm{s}$ in the range of 30--200~pF. 

\begin{figure}[ht]
	\centering
	\includegraphics[height=3.0in,width=3.4in]{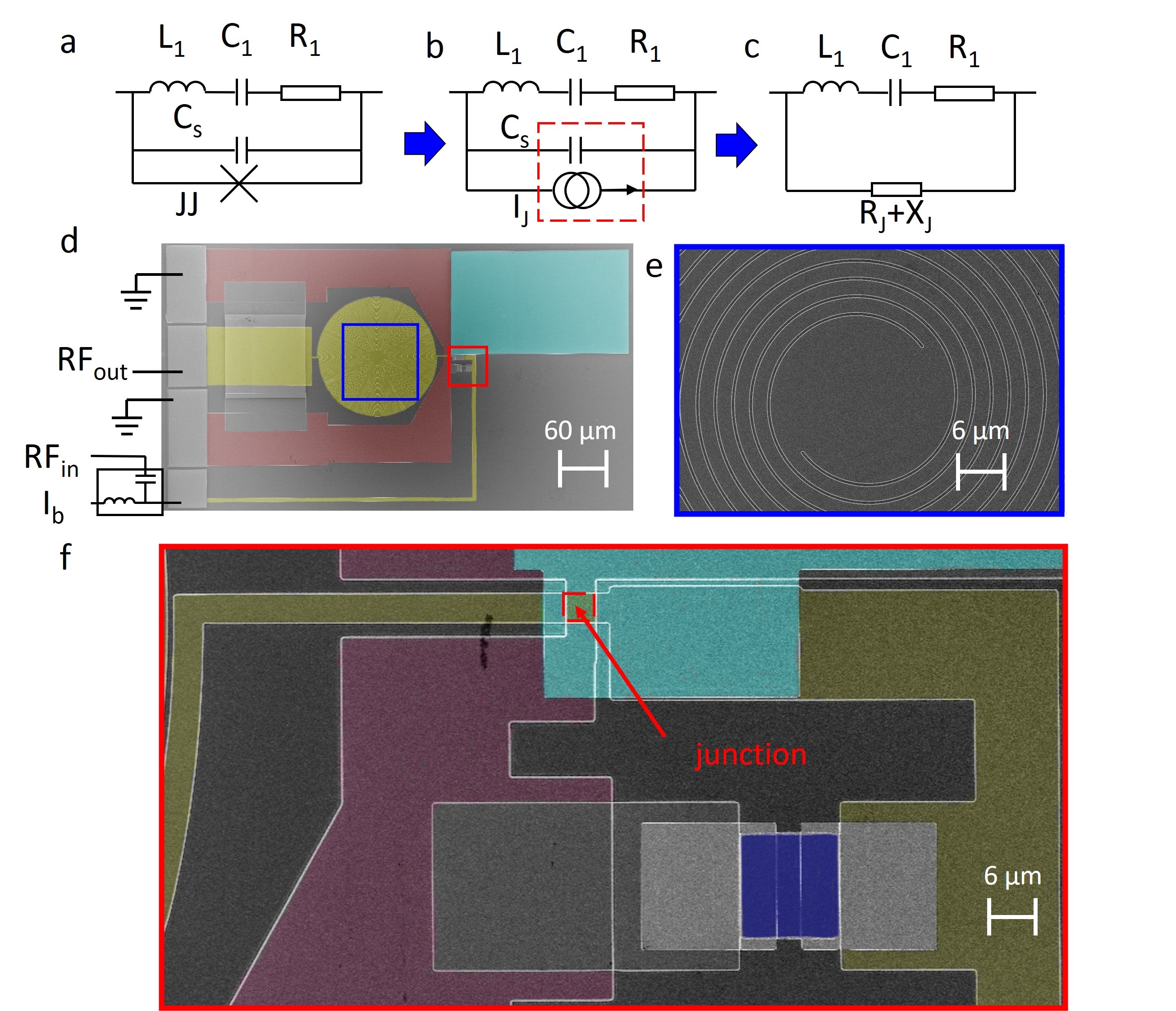}
	
	\caption{\textbf{Working principle and structure of the device.} \textbf{a}, Simplified circuit diagram of the sample where $L_1$ and $C_1$ are the equivalent inductance and capacitance of the microwave resonator, respectively, $R_1$ models the losses of the resonator, and $C_\textrm{s}$ is the shunt capacitor in parallel with the Josepshon junction (JJ). \textbf{b}, In our perturbative treatment (see Methods for mathematical derivation), we fist solve the voltage arising from the microwave current oscillations 
	accounting only for the shunt capacitor and then take into account how this voltage across  the Josephson junction, behaving as a current source,  perturbs the current of the shunt capacitor. \textbf{c}, Resulting impact of the capacitively shunted JJ (red dashed box in (\textbf{a})) on the resonator circuit can be represented by an impedance $R_\textrm{J} + X_\textrm{J}$. A detailed circuit diagram can be found in Fig.~S3. \textbf{d}--\textbf{f}, Overview of the device, consisting of a Josephson junction coupled to a spiral resonator. The ground plane is denoted by pink color. The bias line and the spiral resonator is highlighted in yellow. The shunt capacitor appears in cyan color. The area enclosed by the blue rectangle in (\textbf{d}), is magnified in (\textbf{e}). The red rectangle in (\textbf{d}) denotes the area shown in (\textbf{f}). 
	The junction has a size of $\sim$10 {\textmu}m$^2$. The $\sim$1-$\Omega$ shunt resistor (blue color), with small inductance, functions as an on-chip current to voltage converter for the DC bias. A schematic for the layered structure, which highlights the connections between the different layers, can be found in Fig.~S1. The nominal parameters of the device are designed as follows: the critical current $I_\textrm{c}=10$~{\textmu}A, the shunt capacitance $C_\textrm{s}=192\textrm{ p}\textrm{F}$, and the effective inductance and capacitance of the resonator are $L_{1}=2.0\textrm{ n}\textrm{H}$ and $C_{1}=0.36\textrm{ p}\mathrm{F}$, respectively. 
	}
	
	\label{fig:1}
	
\end{figure}

The utilized harmonic modes in our devices are implemented in planar resonators, either as lumped $LC$ resonators or spiral resonators. In either case, the designed mode profile corresponds to that of the equivalent $L_1$--$C_1$ series resonator as shown in Fig.~\ref{fig:1}a, i.e., there is a current maximum at the junction. The dissipation, including coupling to the load, possible intrinsic damping mechanisms, and the dissipation from the bias circuitry are formally included in the series resistance $R_1$. Note that  the parameters of the equivalent circuit can be extracted by standard circuit analysis from the device design parameters, and hence can be
considered known in the discussion below. Furthermore, we implement the voltage bias across the junction by applying direct current (DC) through a shunt resistor $R_\textrm{s}$ connected in parallel with the junction as shown in Fig.~\ref{fig:1}d. The detailed schemes for implementing the bias and for the coupling of the output radiation are detailed in the Supplementary Information (SI).

Our theoretical approach is based on a perturbative treatment of the C-shunt JJ under a radio-frequency (RF) drive, see Methods section for the detailed derivation. Since $\omega \gg \omega_\textrm{p}$, the voltage bias across the shunted junction is relatively accurately given by the voltage across the shunt capacitor  assuming that all the current drive is simply charging and discharging the capacitor. Thus we define our unperturbed circuit to consist only of the shunt capacitor and the drive. 

In the next step, we treat the Josephson effect as a perturbation that slightly adjusts the current of the shunt capacitor and hence the voltage, see Fig.~\ref{fig:1}b. The gain properties of this driven circuit composing of the junction and the shunt capacitor can be extracted from the Fourier–Bessel expansion of the current-phase relation of a JJ assuming that the junction dynamics is phase-locked to the first Shapiro step with the average voltage of $\left\langle U\right\rangle =\frac{\Phi_{0}\omega}{2\pi}$, where $\Phi_{0}=h/(2e)$. The solution is conveniently represented in Fig.~\ref{fig:1}c by a complex impedance $Z_\textrm{J} = R_\textrm{J} + X_\textrm{J}$, in which $R_\textrm{J}$ and $X_\textrm{J}$ correspond to the real and imaginary components as 
\begin{equation}\label{eq:RJ}
    R_\textrm{J} = -\frac{\hbar\omega\langle I_\textrm{J}\rangle}{eI_{1}^{2}}
\end{equation}
\begin{equation}\label{eq:XJ}
X_\textrm{J} = \frac{1}{\ii\omega C_\textrm{s}}\Bigg(1-\frac{I_\textrm{c}}{I_{1}}[J_{0}(\widetilde i_1)-J_{2}(\widetilde i_1)]\sqrt{1-\Big(\frac{\langle I_\textrm{J}\rangle}{I_\textrm{c}J_{1}(\widetilde i_1)}\Big)^{2}}\Bigg)
\end{equation}

\noindent where $\hbar=h/(2\pi)$, $\langle I_\textrm{J}\rangle$ is the DC current through the junction, $I_1$ is the current amplitude of the self-sustained oscillation of the system at angular frequency $\omega$, $\widetilde i_1 = \frac{2\pi I_{1}}{\Phi_{0}\omega^{2}C_\textrm{s}}$, $I_\textrm{c}$ is the critical current of the junction, and $J_n$ refers to the $n$th-order Bessel function of the first kind. As detailed in the SI, $\langle I_\textrm{J}\rangle$ can be expressed as $\langle I_\textrm{J}\rangle = I_{\mathrm{b}}-\frac{\Phi_{0}\omega}{2\pi R_{s}}$, where $I_\mathrm{b}$ and $R_\mathrm{s}$ are the total bias current and the resistance parallel to the junction at DC, respectively.  The fact that $R_\textrm{J}$ is negative for  positive $\langle I_\textrm{J}\rangle$ is a manifestation of the power generation ability\cite{HGH06}. 

To provide intuitive understanding of the electrodynamics of the source, we note that the power output properties are related to the real component of the impedance, $R_\textrm{J}$, whereas the imaginary component $X_\textrm{J}$ slightly modifies the resonator frequency. In a steady state, the photon emission rate from the JJ to the resonator equals that of the resonator decay, satisfied for $R_\textrm{J} = -R_1$ according to Fig.~\ref{fig:1}c. Formally, we can express the total losses including the Josephson effect by $R_\textrm{J} + R_1$ such that the total effective quality factor assumes the form $Q_\textrm{tot} = \sqrt{L_1/C_1}/(R_\textrm{J}+R_1)$, where we have for simplicity neglected to effect of $X_\textrm{J}$. Thus the sustained oscillations corresponding to $R_\textrm{J} = -R_1$ lead to a diverging $Q_\textrm{tot}$. Due to the amplitude dependence of $R_\textrm{J}$, one obtains at each bias point a single well-defined amplitude for the sustained oscillation. 

The total RF power generation is given by $P_\textrm{out} = -(Q_\textrm{t}/Q_\textrm{e})R_\textrm{J} I_1^2/2 =(Q_\textrm{t}/Q_\textrm{e})\langle I_\textrm{J}\rangle\hbar\omega/(2e)$ where $Q_\textrm{t}$ and $Q_\textrm{e}$ are the experimentally measurable total and loaded quality factors excluding the Josephson effect (see SI). The phase-locking condition implies\cite{KAU81} $\langle I_\textrm{J}\rangle<I_\textrm{c}|J_1(\tilde{i}_{1})|$. Therefore, the maximum power available from the oscillator is $P_\textrm{out,max} =  0.58\times I_\textrm{c}\hbar\omega/(2e)$ where the numeric prefactor follows from the properties of Bessel functions, i.e., from $\max_x{|J_1(x)|}\approx0.58$. Thus, an oscillator for a desired power level and a given frequency may be optimized by choosing $I_\textrm{c}$ accordingly. Technologically, the range of available critical currents spans several orders of magnitude\cite{LMV17}. Note that the maximum power output is obtained by optimizing the load such that $R_1\approx 0.68\frac{\pi I_{\mathrm{C}}}{\Phi_{0}C_{s}^{2}\omega^{3}}$ and minimizing the residual microwave losses such that $\eta_\mathrm{mw} \equiv Q_\mathrm{t}/Q_\mathrm{e}\approx 1$ (see SI). % is to be optimized to obtain the maximum output power with stable phase locking. 
%The related stability criteria and trade-offs in power generation efficiency are analyzed in the SI.

To account for the imaginary component, we note that $X_\textrm{J}$ yields a small correction to the reactance of the excited mode, and results in a small shift in the output frequency. Its bias point dependence maps into the corresponding dependence for the output frequency. This phenomenon is anticipated to affect the quality of the output signal, since a fluctuating bias point thus leads to fluctuations in the frequency and hence in the phase of the output, as studied for different types of Josephson oscillators in the literature\cite{SSY01}. The relative effect of $X_\textrm{J}$ in the total reactance is minimized by maximizing the characteristic impedance of the resonator $\sqrt{L_1/C_1}$. This provides a guideline to resonator optimization in order to achieve higher frequency and phase stability of the output signal: maximize the impedance under the constraints set by the resonator design. Although we can achieve much higher resonator impedances given our design rules, we set here the impedance to roughly 100~$\Omega$ so that the dependence between the emission frequency and the DC bias can be conveniently experimentally observed. This allows us to demonstrate the validity of our model explicitly. 
Further improvement of the linewidth can be realized by increasing the characteristic impedance of the resonator, %which may be obtained by changing the ratio between the center-conductor width and its distance to the ground plane in the coplanar-waveguide design and by using a thin film of high-kinetic-inductance material.    
which may be obtained by decreasing the metallization ratio of the distributed transmission line resonator design and by using a thin film of high-kinetic-inductance material.

%\section{Experiment results}

\vspace*{10pt}
\noindent \textbf{Experimental results}  

\begin{figure}
	\centering
	\includegraphics[height=4.9in,width=3.4in]{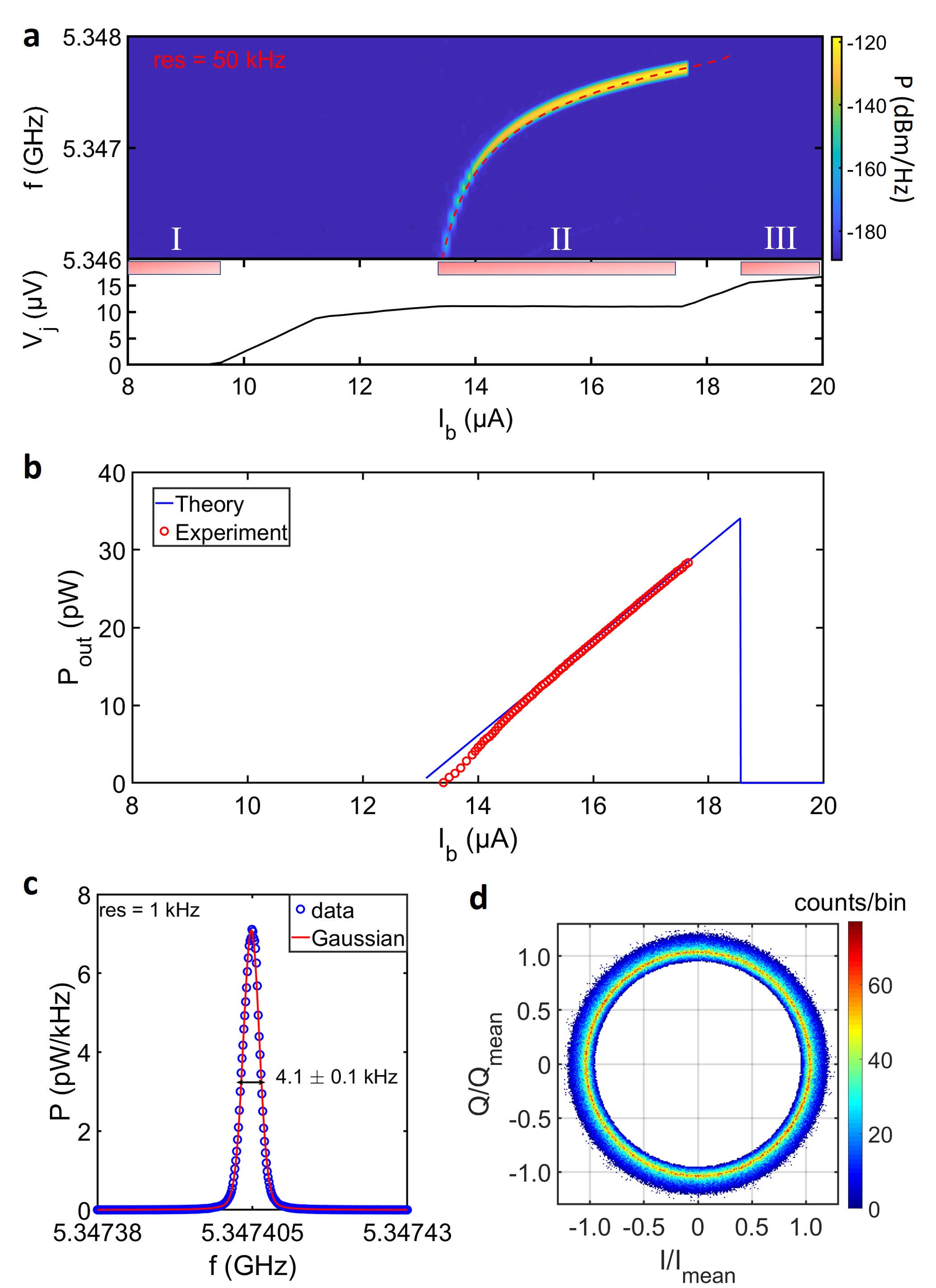}
	
	\caption{\textbf{Characteristics of the spiral-resonator source.} \textbf{a}, Top panel shows the calibrated power spectral density emitted from the source as a function of the bias current and emission frequency. 
	The red dashed trace indicates the position of the emission peak predicted by equation~\eqref{eq:11},  
	where we use only a single adjustable parameter that accounts for the parasitic inductance mainly arising from the bonding wires. The bottom panel shows the direct-current voltage measured across the junction as a function of the bias current. Labels I, II, and III refer to the supercurrent, Shapiro step, and normal-state bias regions, respectively. Results from an extended range can be found in Fig.~S5. \textbf{b}, Measured (red dots) and predicted (blue trace) output power as a function of bias current. \textbf{c}, Line trace of (\textbf{a}) at $I_\textrm{b}$ = 16.5~{\textmu}A (markers) together with a Gaussian fit (solid line). In all figures, 'res' refers to the resolution bandwidth of the spectrum analyzer. \textbf{d}, Measured probability distribution of the source output in the in-phase--quadrature (IQ) plane at $I_\textrm{b}$ = 16.5 {\textmu}A. The data is without digital down-conversion. Here, $10^6$ samples are encoded into 200$\times$200 bins. The obtained ring shape agrees with the coherent nature of the emitted signal. The measurement was carried out at the base temperature of the fridge, $\sim$15~mK. The detailed experiment setup and probability distribution after digital down-conversion can be found in the SI.}
	
	\label{fig:2}
	
\end{figure}

\noindent After the calibration of the gain of the amplification chain (Fig.~S3), the microwave emission spectrum is measured and shown in Fig.~\ref{fig:2}a as a function of the DC bias current $I_\textrm{b}$ which is converted to the DC bias voltage by the on-chip shunt resistor. %MM: Can we give a simple equation for the bias voltage? Is it $R_\textrm{s}(I_\textrm{b}-\langle I_\textrm{J}\rangle)$ like I wrote?
%CY: We get three regions with different behaviors, so perhaps we do not have a single equation that covers the three regions.
We observe three bias regions with distinctive signatures in both the current-voltage (IV) characteristics and the emission spectrum as follows: the supercurrent state for $I_\textrm{b} <$~10~{\textmu}A (region I), the self-induced Shapiro step\cite{BCS99, HGH06} for 13~{\textmu}A~$< I_\textrm{b} <$~18~{\textmu}A (region II), and the normal state with the resistance determined by the shunt resistor for $I_\textrm{b} >$~19~{\textmu}A (region III). There is no photon emission in region I and negligible emission in region III. In the contrast, a major emission occurs at the Shapiro step in region II. The Shapiro step is a manifestation of the self-induced locking of the Josephson and resonator dynamics leading to the measurable power emission.  The central frequency of the emitted signal shifts towards higher frequency with increasing $I_\textrm{b}$ owing to above-mentioned $X_\textrm{J}$ [Fig.~\ref{fig:2}a]. This behaviour is well captured by our model. 

The emitted power, shown in Fig.~\ref{fig:2}b, is obtained by integrating the emission spectrum over frequency. The power increases almost linearly with increasing $I_\textrm{b}$, again well captured by the model. The output power exceeds 20 pW for $I_\textrm{b} >$ 16 {\textmu}A peaking at 28 pW ($-75.5$~dBm) with a corresponding DC power  $P_\textrm{DC}$ = 17.7~{\textmu}A $\times$ 10.7~{\textmu}V = 189 pW ($-67.2$~dBm). This suggests a DC-to-RF conversion efficiency of 15$\%$ at maximum output power.  As shown in Fig.~\ref{fig:2}c, the typical emission linewidth is $4.1\pm 0.1$~kHz 
which is roughly five times as sharp as that obtained in ref.\cite{CBR17} ($\sim$22 kHz). Such a narrow linewidth suggests potential for a noticeable improvement of phase stability over previous coherent cryogenic sources\cite{LST65,AIN07,LSG15,LHS17,LSE17,CLA14,CBR17}.  %The realization of a narrow linewidth without complicated filtering setup in the bias lines also simplifies the experiment setup. 

Interestingly, we use only a single-pole room temperature commercial low-pass filter on the DC line in the present setup, whereas filters at multiple temperature stages have been utilized in previous studies\cite{CBR17}. Thus our filtering scheme relaxes some of the burden required to build and test the experimental setup. However, improvements on the filtering scheme in our setup may lead to further narrowing of the spectral line.

To find evidence that the output of our source is composed of microwaves in a coherent state, we utilize the heterodyne detection technique. The output ﬁeld is demodulated by a local reference tone to yield the in-phase (I) and quadrature-phase (Q) components. The frequency of the reference tone is detuned from the central emission frequency by 62.5 MHz, optimized for our setup. The results of $10^6$ samples are summarized as a two-dimensional (2D) probability distribution depicted in Fig.~\ref{fig:2}d. The probability distribution of the output shows a nearly Gaussian shape with respect to the intensity of the radiation field, or the radius in Fig.~\ref{fig:2}d, as detailed in Fig.~S7. 
The Gaussian ensembles rotates at an intermediate frequency of 62.5 MHz  in the IQ plane, and hence resembles a ring with finite radius and width. This coincides with the distribution of a coherent state averaged over different phase shifts, hence our observation provides evidence on the coherent character of the emission.

%MM: please check all citations to Supplementary figures just before resubmission since those numbers do not update automatically. Also check citations to the panels such as those of Fig. 1 since we have changed the labeling.
In Fig.~S8, we provide data on a reference device differing from that discussed here mainly in its lower critical current, roughly 1.8~{\textmu}A, and having a lumped-element $LC$ resonator. In addition,  the resonator impedance is $\sim\,$3.8~$\Omega$ as opposed to $\sim\,$75~$\Omega$ for the spiral resonator. As expected from the discussion above, the frequency of the emitted signal in Fig.~S8 is much more sensitive to the bias current than for the spiral-resonator sample. Thus the emitted signal is expected to experience excessive phase noise. Nevertheless, the good agreement of the experimental data with our theory for this sample of very different parameters than those of the spiral-resonator sample provides a convenient verification of our model.

To gain more understanding on possible limitations of the linewidth of the generated signal, we utilize the well-established injection locking technique\cite{LSG15, LHS17,CBR17,MDF19}. Here, the frequency of the injection tone $f_\textrm{inj}$ is fixed to that of the free emission at a given bias, whereas the power $P_\textrm{inj}$ is swept.  

\begin{figure}
	\centering
	\includegraphics[height=5.4in,width=3.2in]{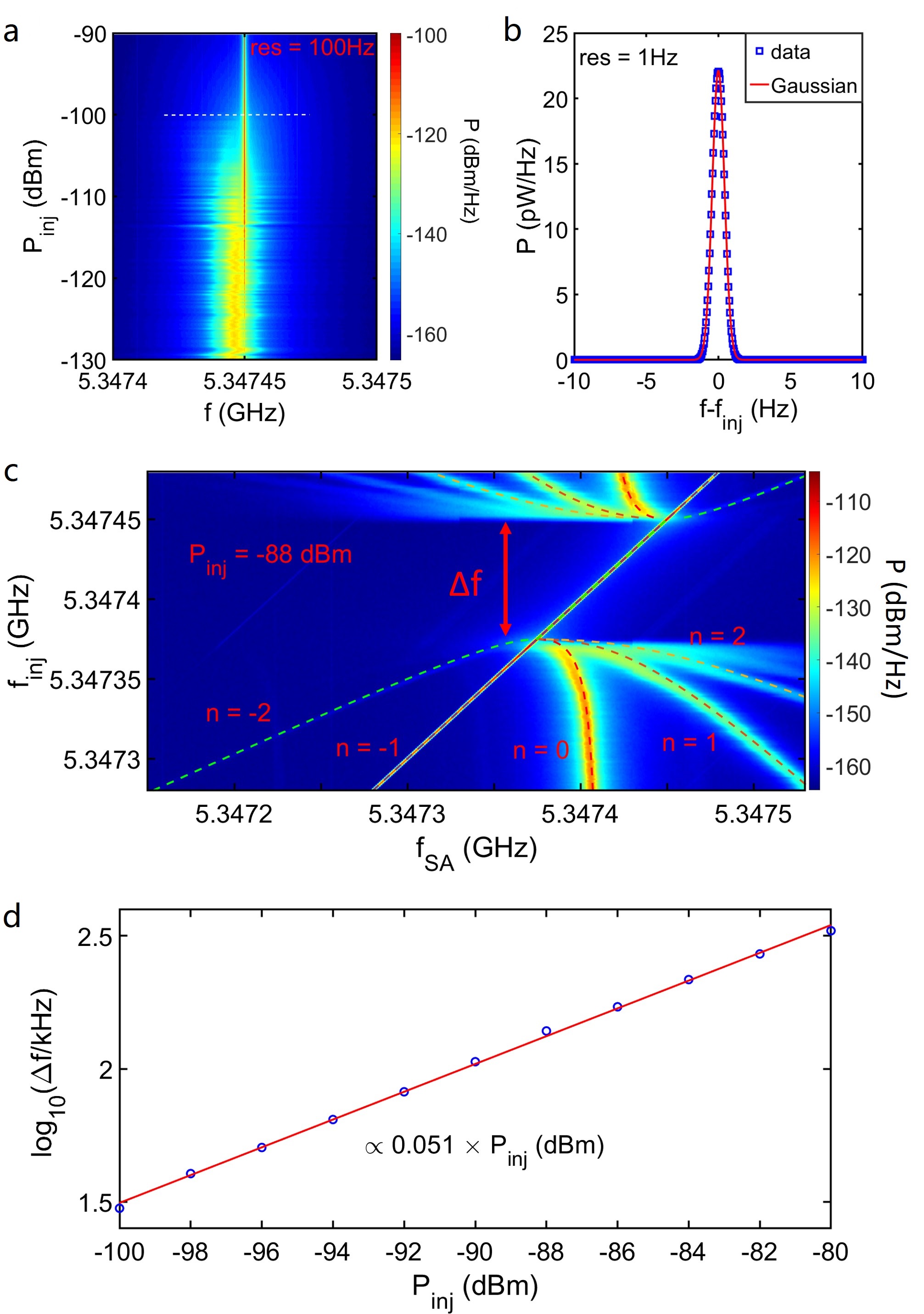}
	
	\caption{\textbf{Emission spectrum with injection locking.} \textbf{a}, Power spectral density of the source emission as a function of injection power $P_\textrm{inj}$ at the sample. The frequency of the injection tone is fixed at 5.34745 GHz. \textbf{b}, Trace of (\textbf{a}) at $P_\textrm{inj} = -100$ dBm (markers) and a Gaussian fit (solid line). Here, the injection tone contributes less than 1$\%$ of the total power. The linewidth of output signal is bounded from below by the minimum resolution of the spectrum analyzer to 1~Hz, which is also responsible for the Gaussian shape. See Fig.~S12 for a more detailed analysis of this data suggesting a linewidth for the output of the source to be of the order of 1~mHz or below. \textbf{c}, Measured power spectral density with respect to the output of the source as a function of the spectral frequency $f_\textrm{SA}$ and of the injection frequency $f_\textrm{inj}$ at a fixed injection power of -88 dBm. $\Delta f$ is defined as the width of the frequency range where the emission signal is phase locked to the injection tone. Here $P_\textrm{inj} = -88$ is chosen to highlight bending of peaks under injection locking condition. Similar plots at other injection power can be found in SI. \textbf{d}, The frequency range $\Delta f$ as a function of the injection power $P_\textrm{inj}$. The measured data (blue circles) are in good agreement with the Adler theory (solid line). Note that the horizontal axis is in the units of power, whereas in the case of Adler theory, it is often in the units of voltage.}
	
	\label{fig:3}
	
\end{figure}

Figure~\ref{fig:3}a illustrates that in our experiments, the injection tone induces a very sharp peak into the emission spectrum, into which the whole emission gradually shifts with increasing injection power. For $P_\textrm{inj} > -100$ dBm, we observe only a single emission peak with a linewidth of 1~Hz (Fig~\ref{fig:3}b). Interestingly, this linewidth is limited by the smallest resolution bandwidth of the used spectrum analyzer and our more detailed study of the measured spectrum shown in Fig.~S12 suggests that the linewidth of the injection-locked source is of the order of 1~mHz or below. It is also possible to measure very small linewidths with an advanced hardware setup, for example, by carrying out a Fourier transform of the IQ traces after sufficient averaging. Yet, we note that, the typical linewidths of state-of-the-art superconducting qubits are in the kilohertz range, i.e., comparable to the linewidth of our source without injection locking, see Fig.~\ref{fig:2}c. We have also measured the emission spectrum  with a ﬁxed $P_\textrm{inj}$ but varying $f_\textrm{inj}$, and the results agree well with Adler theory, as shown in Figs.~\ref{fig:3}c and~d, and discussed in the SI.

\begin{figure}
	\centering
	\includegraphics[height=3.8in,width=3.4in]{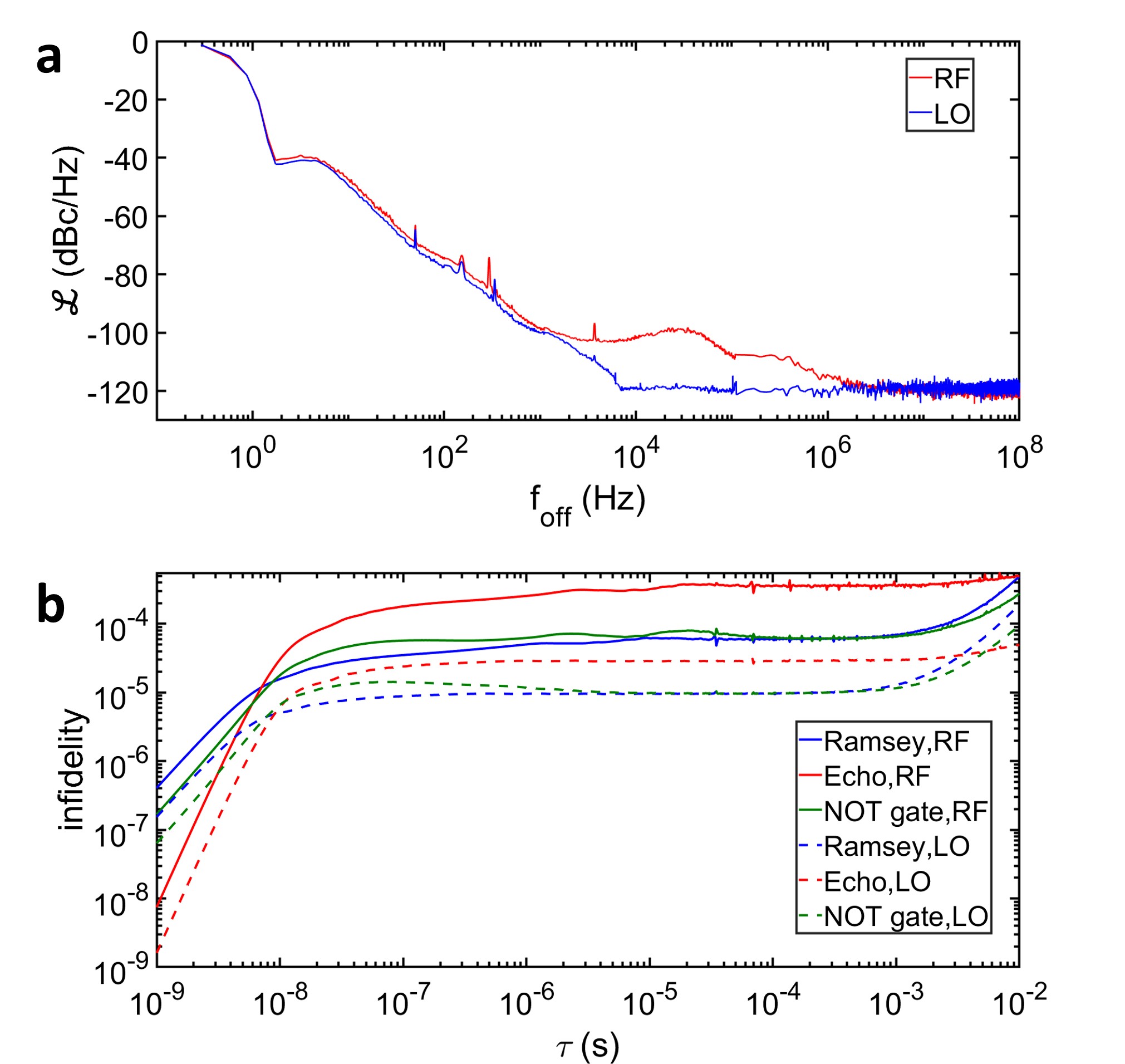}
	
	\caption{\textbf{Noise properties of the on-chip source.} \textbf{a}, Single-sideband phase noise $\mathscr{L}$ of the emitted signal (red line) and of the local oscillator only (blue line) as a function of the offset frequency with respect to the injection tone frequency. The on-chip source is injection-locked at $P_\textrm{inj} = -100$ dBm. In the measurement with only the local oscillator, we adjusted its power to $\sim -77$~dBm, i.e., equal to the RF signal power emitted by the on-chip source, for the direct comparison of phase noise. 
	\textbf{b}, Inﬁdelities of Ramsey (blue color), Hahn echo (red color) and NOT gate (green color) operations arising from the on-chip source (solid lines) and the local oscillator (dashed lines) as functions of the evolution time $\tau$. For the Ramsey and echo operations, $\tau$ refers to the total duration of free evolution. For the NOT gate, $\tau$ is the duration of the $\pi$ pulse.} 
	
	\label{fig:4}

\end{figure}

We extract the the phase noise $\mathscr{L}$ from the emission spectrum under injection locking with $P_\textrm{inj} = -100$ dBm where the injection tone contributes less than $1\%$ of the total power. Our results presented in Fig.~\ref{fig:4}a show that $\mathscr{L}$ decays rapidly with increasing frequency offset $f_\textrm{off}$ from $f_\textrm{inj}$. It reaches $-95$~dBc/Hz at $f_\textrm{off}=10$~kHz, which is about 15~dB below the corresponding value for a typical lab-grade local oscillator (LO) operating at room temperature\cite{BOB16}, but it needs further improvement to be compatible with high-precision LOs such as that used to generate the injection tone. The measured phase noise eventually saturates to $-120$~dBc/Hz at $\sim$5 MHz. The saturation is mainly determined by noise added by our amplification chain (Fig.~S10). The noise floor can be possibly subtracted to a large extent from the source noise by averaging the data carefully for the offset frequency exceeding 5~MHz (Fig.~\ref{fig:4}).

It is possible to minimize the influence of the system noise using the cross-correlation technique\cite{WWF92} and thus to obtain the actual $\mathscr{L}$ at large $f_\textrm{off}$. We leave this extension for future research. Nevertheless, we note that $\mathscr{L} = -116$~dBc/Hz at an offset frequency of 1~MHz is well below $-99$~dBc/Hz obtained by the quantum-dot-based on-chip microwave source studied in ref.\cite{LHS17}.  

Owing to the large output power and hence the large signal-to-noise ratio, the phase noise $\mathscr{L}$ yields the dominating noise of the device up to relatively large offset frequencies. This motivates us to examine the influence of the phase noise on apparent qubit dephasing and on the gate and operation ﬁdelity. We consider the source, augmented with a noiseless pulse and phase modulator, to drive an ideal qubit that is free of intrinsic dephasing and dissipation. Following the framework of ref.\cite{BOB16}, we calculate the infidelity of the quantum operations, defined as $1 - F_\textrm{av}(\tau)$, where the averaged operational fidelity is denoted by $F_\textrm{av}(\tau)$. We have 
$$F_\textrm{av}(\tau) \approx \frac{1}{2} [1+\textrm{e}^{-X(\tau)}]$$ 
where the evolution time is denoted by $\tau$, 
$$X(\tau)=\frac{1}{2\pi}\int_{0}^{\infty}\textrm{d}f\,10^{\mathscr{L}(f)/(10\textrm{ dBc/Hz})}\frac{1}{\textrm{Hz}}  \sum_{l \in x,y,z}G_{z,l}(f,\tau)$$
and $G_{z,l}(f,\tau)$ is a filter function that quantifies the action of the control Hamiltonian %Here, $\mathscr{L}$ is expressed in the unit of dBc/Hz. 
and hence depends on the specific quantum operation as discussed in the SI.

Figure~\ref{fig:4}b shows the calculated infidelity for prototypical quantum operations: Ramsey,  Hahn echo, and NOT gate operations. The infidelity is $\sim\,$0.1$\%$ for all these operations after a long evolution time of $\tau=10$~ms. These infidelities are an order of magnitude lower than those achieved by a typical lab-grade LO\cite{BOB16}. However, in the short $\tau$ limit, the calculated infidelity is about an order of magnitude larger than that obtained from lab-grade LO\cite{BOB16} due to the overestimated phase noise $\mathscr{L}$ at large $f_\textrm{off}$ arising from the amplification chain. The low-offset-frequency components dominate in the long $\tau$ limit. On the other hand, both low- and high-frequency components have a noticeable contribution in the short $\tau$ limit as discussed in the SI. For comparison, the infidelities of the operations are reduced by an order of magnitude if extract them from the phase noise measured for the LO. These infidelities are also significantly affected by the noise floor set by the amplification chain. Same calculation with the noise floor subtracted can be found in the SI.

The above-measured low noise of our microwave source suggests that the device is suitable for controlling state-of-the-art superconducting qubits with coherence times currently reaching 100~{\textmu}s\cite{RGP12, RBM19, XHC20}. \\

\clearpage
\noindent \textbf{Conclusions} 

\noindent In this work, we demonstrated an on-chip coherent microwave source realized by a Josephson junction strongly coupled to a spiral resonator. The source can generate microwave signals with narrow linewidth ($<1$~Hz, suggestively $\lesssim 1$~mHz), low noise ($<-120$~dBc/Hz), high output power ($>25$~pW), %MM: Why 25 pW here? Why not 28 pW? 
%CY:28 pW is the peak value. 25 pW may be thought as a typical value.
and a fine dc-to-rf power conversion efficiency of about 15\%. The output power is two orders of magnitude higher than that previously reported for double-quantum-dot sources\cite{LSG15, LHS17, LSE17} or an aluminum-junction source\cite{CBR17}. We confirm that the expected infidelity bound arising from the phase noise of our source in a typical quantum-logic operation is below 0.1$\%$ up to 10-ms evolution ensuring that the signal quality is sufficient for the control of state-of-the-art quantum systems.  

We used the injection-locking technique in an effort to study the intrinsic limitations of the oscillator. As for any oscillating source independent of the technology, frequency and phase stabilization techniques need to be used. An alternative to this injection locking scheme is to bring the reference tone to the cryogenic temperature at a low-frequency band restricting the bandwidth requirements from room temperature, and to use frequency multiplication techniques\cite{HSD16} to generate the injection tone. Furthermore, integration of superconducting quantum interference devices (SQUIDs) with the source allows to tune the emission frequency without significant amplitude modulation, and hence enables frequency stabilization based on phase-locked loops which are typically used in the context of voltage-controlled oscillators at room temperature, and for which superconducting counterparts have been demonstrated as well\cite{VSP02}.

In general, there are several approaches pursued in the field of integrated control of quantum systems, including cryogenic semiconductor-based techniques, or even those based on optical-to-microwave transducers\cite{GMS20}.  Semiconductor-based oscillators have been demonstrated with the output power of about 0.2 {\textmu}W at 1.5~K\cite{ANG19}. Full semiconductor-based cryogenic control systems have been demonstrated at the operating temperatures of a few kelvin, with power consumption of the order of 100~mW\cite{BPS20}. Cryogenic semiconducting electronics 
has been demonstrated at the tempearture of 100 mK, with power consumption of 18~nW per cell\cite{PDK21}. While less matured, all-superconducting control electronics concepts are likely superior in power efficiency. 
Ideally, the power dissipated at the base temperature is efficiently converted into the control signals of the quantum system. An order-of-magnitude estimate for our source, assuming a 100-ns-long control period or readout pulses of 10 photons at 5.3 GHz enables driving about 7000 qubits with the output power available. 

The power and signal quality alone does not yet guarantee the source to be useful in the control of quantum systems. A feasible scheme for millikelvin-operated full waveform control may be a combination of our source with cryogenic microwave phase shifters\cite{KOL17, ZLK20} or flux-tunable resonators\cite{OSS07} and quantum-circuit refrigerators\cite{TPL17, SMS19}, enabling power-efficient waveform control. Moreover, cryogenic mixing or parametric frequency conversion techniques are an option for full waveform generation. 

Importantly, we anticipate that a stable microwave source is useful in other contexts than direct qubit control such as those based on single-flux-quantum (SFQ) logic\cite{LMV19} needing a master clock reference. Compared to semiconductor counterparts, SFQ is orders of magnitude more power efficient\cite{SPH06}.  In general, it is understood that the field of integrated control systems in cryogenic solid-state quantum technology is still at its infancy. Independent of the detailed realization, it is expected that a power-efficient low-phase-noise reference oscillator is a necessary component, acting as the primary source of microwave power and as reference master clock. 

%MM: I wrote below about the scalability and thermal noise. We could update the response letter accordingly
In the future, we aim to study the properties of cascaded cryogenic sources, where one injection-locked source works as the  locking tone for other sources. In such a scenario, the total injection power delivered from room temperature, and hence the required amount of microwave cables, is independent of size of the cryogenic control system such as the number of qubits in a large-scale quantum computer. 

Furthermore, we aim to study the thermalization of the output impedance of the source. Although the measured phase noise also includes the effect of finite temperature, thermal photons leaking to different parts of a quantum computer may lead to additional undesired dephasing. Fortunately, our preliminary thermal analysis and that in ref.\cite{YHZ19} suggests that the shunt resistor can be thermalized well below 100~mK even at high output powers. In fact, a cryogenic source may also be able to utilize the exponential thermal suppression of noise photons at the signal frequency, in stark contrast to signals generated at room temperature, after which the suppression of noise photons by cryogenic attenuation or filtering leads to equal suppression of the signal power.

\clearpage
\noindent \textbf{Methods}

\noindent\textbf{Theoretical model.} Let us develop an analytic model for the Josephson oscillators based on a capacitively shunted Josephson junction coupled in parallel with a resonator circuit. The concept is to first derive an analytic model for the junction as a gain element. Then the gain properties are analyzed for the resonator-coupled junction. The mathematical tools used include (i) well-known conditions of junction phase-locking to an RF drive, (ii) a perturbative harmonic analysis of the junction to provide the gain properties under the RF drive and the phase-locking conditions, and (iii) showing by a power-balance criterion that the gain and phase-locking conditions satisfy a stable sustained oscillating mode by the self-generated RF drive. For (ii), the junction properties are described as an effective impedance, the negative real part of which is the manifestation of the gain. The model provides a convenient engineering tool for obtaining the basic properties of the oscillator. Namely, it provides simple design criteria for stable oscillator operation that can be used to predict the output power, DC-to-RF conversion efficiency, and curiously, the operation-point-dependent output frequency.  

We begin our analysis from an RF-driven capacitively shunted Josephson junction, as shown in Fig~\ref{fig:1}. In the first step, let us consider a bare capacitively shunted Josephson junction. Assume that the junction is subjected to an RF current $I_\textrm{RF}=I_{1}\sin(\omega t)$ such that $\omega\gg\omega_\textrm{p}$, where $\omega_\textrm{p}=1/\sqrt{L_\textrm{J}C_\textrm{s}}$, is the junction plasma frequency, $L_\textrm{J}$ is the Josephson inductance, and $C_\textrm{s}$ is the capacitance parallel to the junction. The capacitance $C_\textrm{s}$ is dominated by the shunt capacitance since the intrinsic capacitance of the junction is negligible. Under the above assumptions, the RF drive couples predominantly capacitively through the parallel connection of the junction and the shunt capacitance. 
Thus, the unperturbed voltage, neglecting the Josephson effect, across the tunnel junction is simply
\begin{equation}\label{eq:1}
V_\textrm{T}(t)= \frac{1}{C_\textrm{s}}\int I_{1}\sin(\omega t)\textrm{d}t=-\frac{I_{1}}{\omega C_\textrm{s}}\cos(\omega t)+\langle U\rangle%MM: Taking out commas and periods after online equations as per Nature style. Check.  
\end{equation}
where the constant of integration is the DC voltage across the junction. 

We employ the AC Josephson relation to obtain the phase across the junction as
\begin{equation}\label{eq:2}
\phi=\frac{2\pi}{\Phi_{0}}\int V_\textrm{T}\,\textrm{d}t=-\frac{2\pi I_{1}}{\Phi_{0}\omega^{2}C_\textrm{s}}\sin(\omega t)+\frac{2\pi}{\Phi_{0}}\langle U\rangle t-\phi_\textrm{c}
\end{equation}
where we have defined the constant of integration $-\phi_\textrm{c}$. Next, we utilize the Josephson current-phase relation $ I_\textrm{J}=I_\textrm{c}\sin\phi$ which provides the current through the Josephson tunnel element as
\begin{equation}\label{eq:3}
I_\textrm{J}=I_\textrm{c}\sin[-\widetilde i_1 \sin(\omega t)+\omega t-\phi_\textrm{c}] 
\end{equation}
where we have defined $\widetilde i_1 = \frac{2\pi I_{1}}{\Phi_{0}\omega^{2}C_\textrm{s}}$ and written $\langle U\rangle=\Phi_{0}\omega/(2\pi)$ by adopting the phase-locking condition, i.e., the system is biased on the first Shapiro step. This condition is analyzed in the SI %MM: Check where
to check the validity range of the solutions. Here, we merely assume it. We rewrite equation~\eqref{eq:3} with the help of its Fourier–Bessel series as
\begin{equation}\label{eq:4}
I_\textrm{J}(I_{0}, \ I_{1},\ t)=I_\textrm{c}\sum_{m=-\infty}^{\infty}J_{m}(\widetilde i_1)\sin[\omega t(1-m)-\phi_\textrm{c}]
\end{equation}

Let us first consider the direct current $\langle I_\textrm{J}\rangle$ through the tunnel element. In equation~\eqref{eq:4} this follows from the term $m=1$ as
\begin{equation}
\langle I_\textrm{J}\rangle=I_\textrm{c}J_{1}(\widetilde i_1)\sin(-\phi_\textrm{c})
\end{equation}
In fact, $\langle I_\textrm{J}\rangle = I_{\mathrm{b}}-\frac{\Phi_{0}\omega}{2\pi R_{s}}$ is related to external direct bias current (see SI) $I_{\mathrm{b}}$.
Solving for $\phi_\textrm{c}$ yields
\begin{equation}\label{eq:6}
 \phi_\textrm{c}=\arcsin\Big(-\frac{\langle I_\textrm{J}\rangle}{I_\textrm{c}J_{1}(\widetilde i_1)}\Big)
\end{equation}

To address the RF properties, we consider the fundamental-frequency component of $I_\textrm{J}(t)$ which follows from equation~\eqref{eq:4}, picking terms $m=0$ and $m=2$ as 
\begin{equation}
I_{J1}(t)=I_\textrm{c}[J_{0}(\widetilde i_1)\sin(\omega t-\phi_\textrm{c})+J_{2}(\widetilde i_1)  \sin (-\omega t-\phi_\textrm{c})]
\end{equation}
Using a Bessel recurrence formula and equation~\eqref{eq:6}, leads to
\begin{equation}
I_{J1}(t)= I_\textrm{c}[J_{0}(\widetilde i_1)-J_{2}(\widetilde i_1)]\sqrt{1-\Big(\frac{\langle I_\textrm{J}\rangle}{I_\textrm{c}J_{1}(\widetilde i_1)}\Big)^{2}}\sin(\omega t)+\frac{2\langle I_\textrm{J}\rangle}{\widetilde i_1J_{1}(\widetilde i_1)}J_{1}(\widetilde i_1)\cos(\omega t)
\end{equation}

In the capacitively shunted junction case, the RF current $I_{J1}$ couples back to the shunt capacitance since we assume that it is the dominant impedance at frequency $\omega$. As this happens, a voltage emerges on top of $V_\textrm{T}$ of equation~\eqref{eq:1}. Marking this voltage perturbation as $V_\textrm{P}(t)$ we obtain 

\begin{equation}\label{eq:9}
V_\textrm{P}(t)=\frac{I_\textrm{c}}{\omega C_\textrm{s}}[J_{0}(\widetilde i_1)-J_{2}(\widetilde i_1)]\sqrt{1-\Big(\frac{\langle I_\textrm{J}\rangle}{I_\textrm{c}J_{1}(\widetilde i_1)}\Big)^{2}}\cos(\omega t)-\frac{2\langle I_\textrm{J}\rangle}{\widetilde i_1\omega C_\textrm{s}}\sin(\omega t)
\end{equation}

Let us convert equations~\eqref{eq:1} and~\eqref{eq:9} into frequency plane by identifying the in-phase $\sin(\omega t)$ and quadrature $\cos(\omega t)$ components to write $\widetilde V_\textrm{tot}(\omega)=\widetilde V_\textrm{T}(\omega)+\widetilde V_\textrm{P}(\omega)$. Furthermore, it is practical to write the output in the form of impedance $Z_\textrm{J}(\omega)=\widetilde V_\textrm{tot}(\omega)/I_{1}$. From equations~\eqref{eq:1} and~\eqref{eq:9} and by arranging the quadratures, it follows
\begin{equation}\label{eq:10}
Z_\textrm{J}(\omega)=  -\frac{2\langle I_\textrm{J}\rangle}{\widetilde i_1 I_{1}\omega C_\textrm{s}}-\ii\frac{1}{\omega C_\textrm{s}}+\ii\frac{I_\textrm{c}}{\omega C_\textrm{s}I_{1}}[J_{0}(\widetilde i_1)-J_{2}(\widetilde i_1)]\sqrt{1-\Big(\frac{\langle I_\textrm{J}\rangle}{I_\textrm{c}J_{1}(\widetilde i_1)}\Big)^{2}}
\end{equation}
Insertion of the definition $\widetilde i_1 = \frac{2\pi I_{1}}{\Phi_{0}\omega^{2}C_\textrm{s}}$ here yields
\begin{equation}\label{eq:11}
Z_\textrm{J}(\omega)= -\frac{\omega\Phi_{0}\langle I_\textrm{J}\rangle}{\pi I_{1}^{2}} +\frac{1}{\ii\omega C_\textrm{s}}\Bigg\{1-\frac{I_\textrm{c}}{I_{1}}[J_{0}(\widetilde i_1)-J_{2}(\widetilde i_1)]\sqrt{1-\Big(\frac{\langle I_\textrm{J}\rangle}{I_\textrm{c}J_{1}(\widetilde i_1)}\Big)^{2}}\Bigg\}
\end{equation}
We stress that the real part of the junction impedance is negative if $\langle I_\textrm{J}\rangle>0$, and hence represents gain. Note that here the direction of positive current is fixed by the choice that we operate at the first positive-voltage Shapiro step. The imaginary part of the junction impedance equals to that of the shunt capacitor modified by a nontrivial perturbative term due to the Josephson effect. Equation~\eqref{eq:11} yields equations~\eqref{eq:RJ} and~\eqref{eq:XJ} for the real and the imaginary component of the junction impedance, respectively.

\vspace*{10pt}
\noindent\textbf{Device fabrication.} The devices are fabricated 
in a multi-layer process for superconducting circuits, the key element of which is a sidewall-passivated-spacer (SWAPS) technique\cite{LMV17} for the Nb-Al/AlO${}_x$-Nb Josephson junctions. Figures~1b and~S1 summarize the structure of the device. The fabrication begins with a hydrofluoric-acid (HF) dip of a 150-mm high-resistivity silicon wafer to remove oxides from the surface. The trilayer stack for the junctions with thicknesses 100~nm/10~nm/100~nm is then deposited, with a target critical-current density of 100~A/cm${}^2$. We deposit the subsequent layers as follows: a second 120-nm-thick niobium layer, an atomic-layer-deposited 50-nm AlO${}_x$ insulator for a parallel-plate capacitance density of roughly 1.5~fF/\textmu$\text{m}^2$, a third niobium layer of 120-nm thickness, and a normal-metal 
layer with thickness of about 100--130~nm for a sheet resistance of approximately 4~$\Omega/\square$. The patterning of the layers is enabled by ultraviolet (UV) photolithography. The niobium layers are etched with plasma, whereas the insulators and the resistors are wet etched. The Josephson oscillators are fabricated in the same batch of wafers as the traveling-wave parametric amplifiers of ref.\cite{SVM20}.\\

%\clearpage
%\bibliography{main}% Produces the bibliography via BibTeX.

\providecommand{\noopsort}[1]{}\providecommand{\singleletter}[1]{#1}%

\vspace*{10pt}
\noindent \textbf{Acknowledgements} 

\noindent We have received funding from the Centre for Quantum Engineering at Aalto under Grant number JVSG; European  Research Council under Grant No~681311 (QUESS) and MarieSkłodowska-Curie Grant No.~795159; Academy of Finland through its Centres of Excellence Programme (project numbers ~312300, ~312059 and ~312295) and grants (numbers ~314447, ~314448, ~314449, ~305237, ~316551, ~308161, ~335460, and ~314302). We thank Hannu Sipola, Roope Kokkoniemi, Jean-Philippe Girard, and Jukka Pekola for useful discussion.\\

\clearpage 

\clearpage

\clearpage

\clearpage

\clearpage 

\begin{table*}
\centering
\caption{\textbf{Comparison of several typical low temperature oscillators.} Key parameters of cryogenic sources gathered from the indicated references. Here, 'N/A' refers to a case where the corresponding data was not found.}
\label{table:1}
\begin{tabular}{|p{50mm}|p{25mm}|p{15mm}|p{18mm}|p{40mm}|} 
 \hline
 Device & Operation Temperature & Output Power & Linewidth & Phase Noise \\ 
 \hline
 Nb junction device (this work) & 10 mK & 28 pW & 4 kHz & -116 dBc/Hz at 1 MHz \\ 
 \hline
 Al junction device\cite{CBR17} & 10 mK & 0.255 pW & 22 kHz & N/A \\ 
 \hline
 Double-quantum-dot\cite{LHS17} & 10 mK & 0.2 pW & 5.6 kHz & -99 dBc/Hz at 1.3 MHz \\ 
 \hline
 SiGe HBT oscillator\cite{HJK18} & 4 K & 0.2 {\textmu}W & 200 kHz & -112 dBc/Hz at 1 MHz \\ 
 \hline
 Cryogenic HEMT oscillator\cite{ANG19} & 1.4 K & 0.2 {\textmu}W & N/A & -112 dBc/Hz at 1 MHz \\ 
 \hline
\end{tabular}
\end{table*}

\vspace*{10pt}

\end{document}